\documentclass[9pt]{article}
\usepackage{spconf,amsmath,graphicx,url}

\def\L{{\cal L}}

\title{Transformer based unsupervised pre-training for acoustic representation learning}
\name{\begin{tabular}{c}Ruixiong Zhang, Haiwei Wu, Wubo Li, Dongwei Jiang, Wei Zou, Xiangang Li\end{tabular}}
\address{DiDi Chuxing, Beijing, China}

\begin{document}
%
\maketitle
\begin{abstract}
  Recently, a variety of acoustic tasks and related applications arised. For many acoustic tasks, the labeled data size may be limited. To handle this problem, we propose an unsupervised pre-training method using Transformer based encoder to learn a general and robust high-level representation for all acoustic tasks. Experiments have been conducted on three kinds of acoustic tasks: speech emotion recognition, sound event detection and speech translation. All the experiments have shown that pre-training using its own training data can significantly improve the performance. With a larger pre-training data combining MuST-C, Librispeech and ESC-US datasets, for speech emotion recognition, the UAR can further improve absolutely 4.3\% on IEMOCAP dataset. For sound event detection, the F1 score can further improve absolutely 1.5\% on DCASE2018 task5 development set and 2.1\% on evaluation set. For speech translation, the BLEU score can further improve relatively 12.2\% on En-De dataset and 8.4\% on En-Fr dataset.
\end{abstract}
\section{Introduction}
\label{sec:intro}
The goal of acoustic representation learning is to transform raw or surface features into the high-level feature which are more accessible to acoustic tasks\cite{tzanetakis2000marsyas}. It is critical to make acoustic representations more general and robust to improve the performance of acoustic tasks. However, the labeled data size of the specific acoustic task may be limited so that the learned representations can be less robust and the performance can be vulnerable to unseen data. On the other hand, there exists varieties of acoustic tasks which range from speaker verification, speech recognition to event and scene detection. For supervised learning, the learned representation useful for one task may be less suited for another task. It is worthwhile to explore how to utilize all kinds of datasets to learn a general and robust representation for all kinds of acoustic tasks.

Unsupervised pre-training can provide an appealing method to learn more general and robust high-level features that are less specialized towards solving a single supervised task. The training objective of unsupervised pre-training is only related with acoustic features themselves and is not dependent on any other downstream target. Because of this advantage, much more unlabeled data can be utilized so that a larger and more general model can be learned. At the same time, the learned representations can be directly utilized or fine-tuned for specific downstream tasks. 

Contrastive Predictive Coding(CPC)\cite{oord2018representation} has provided a universal unsupervised learning approach to extract useful representations from high-dimensional data. The autoregressive mechanism is used for predicting future information. However, it can only be applied in uni-directional models. Masked Predictive Coding(MPC)\cite{jiang2019improving} has been proposed to utilize speech data in an unsupervised manner for speech recognition. It uses the bidirectional transformer based architecture and uses Masked-LM\cite{devlin2018bert} like structure to perform predictive coding. The pre-trained representations can be further fine-tuned to improve specific speech recognition tasks. However, the speech or acoustic representation pre-trained from this method has not yet been applied to other kinds of acoustic tasks and also the performance of this unsupervised pre-training method on non-speech audio tasks remains unknown. 

In this paper, we get intuition from MPC and utilize a Transformer\cite{vaswani2017attention} based unsupervised pre-training method for acoustic representation learning. Transformer based encoder can be pre-trained by a large amount of unlabeled audio from various kinds of datasets. After pre-training, all we should do is to add a decoder layer targeted for downstream tasks and fine-tune the whole model. we have demonstrated that our method can learn a more general and robust acoustic representation which can significantly improve the performance of various kinds of acoustic tasks.

\section{Related Work}
\label{sec:related work}
Contrastive Predictive Coding(CPC) provided a universal unsupervised learning approach and the learned representation is able to achieve strong performance on four domains: speech, images, text and reinforcement learning in 3D environments. This model is mainly composed of two parts: a non-linear encoder $g_{enc}$ and an autoregressive model $g_{ar}$ . Given an input sequence $(x_{1} , x_{2} , . . . , x_{T})$, $g_{enc}$ encodes observations $x_{t}$ to a latent embedding space $z_{t} = g_{enc}(x_{t})$ and $g_{ar}$ accepts $z_{t}$ to produce a context representation $c_{t} = g_{ar}(z_{\leq t})$. Targeting at predicting future observations $x_{t+k}$ ,a density ratio $f (x_{t+k} , c_{t} )$ is modelled to maximally preserve the mutual information between $x_{t+k}$ and $c_{t}$. 
To optimize $g_{enc}$ and $g_{ar}$ , the contrastive loss is minimized:
\begin{equation}
	\L_N = -\displaystyle \mathop{{E}}_{X}[log\frac{f(x_{t + k}, c_t)}{\sum_{x_j\in X}f_k(x_j, c_t)}],
\end{equation}
where $N$ represents number of samples in $X = {x_{1} , x_{2} , . . . , x_{N}}$, with one positive sample from distribution $p(x_{t+k} |c_{t} )$ and the rest being negative samples from distribution $p(x_{t+k})$.

Autoregressive Predictive Coding(APC)\cite{chung2019unsupervised} also proposed an autoregressive model for unsupervised speech representation learning. It used a deep LSTM network and make the model to predict further steps ahead of the current frame during training. APCs have demonstrated a strong capability of extracting useful phone and speaker information.
\begin{figure*}[htb]
	\begin{minipage}[b]{1.0\linewidth}
		\centering
		\centerline{\includegraphics[width=17.0cm,height=9.5cm]{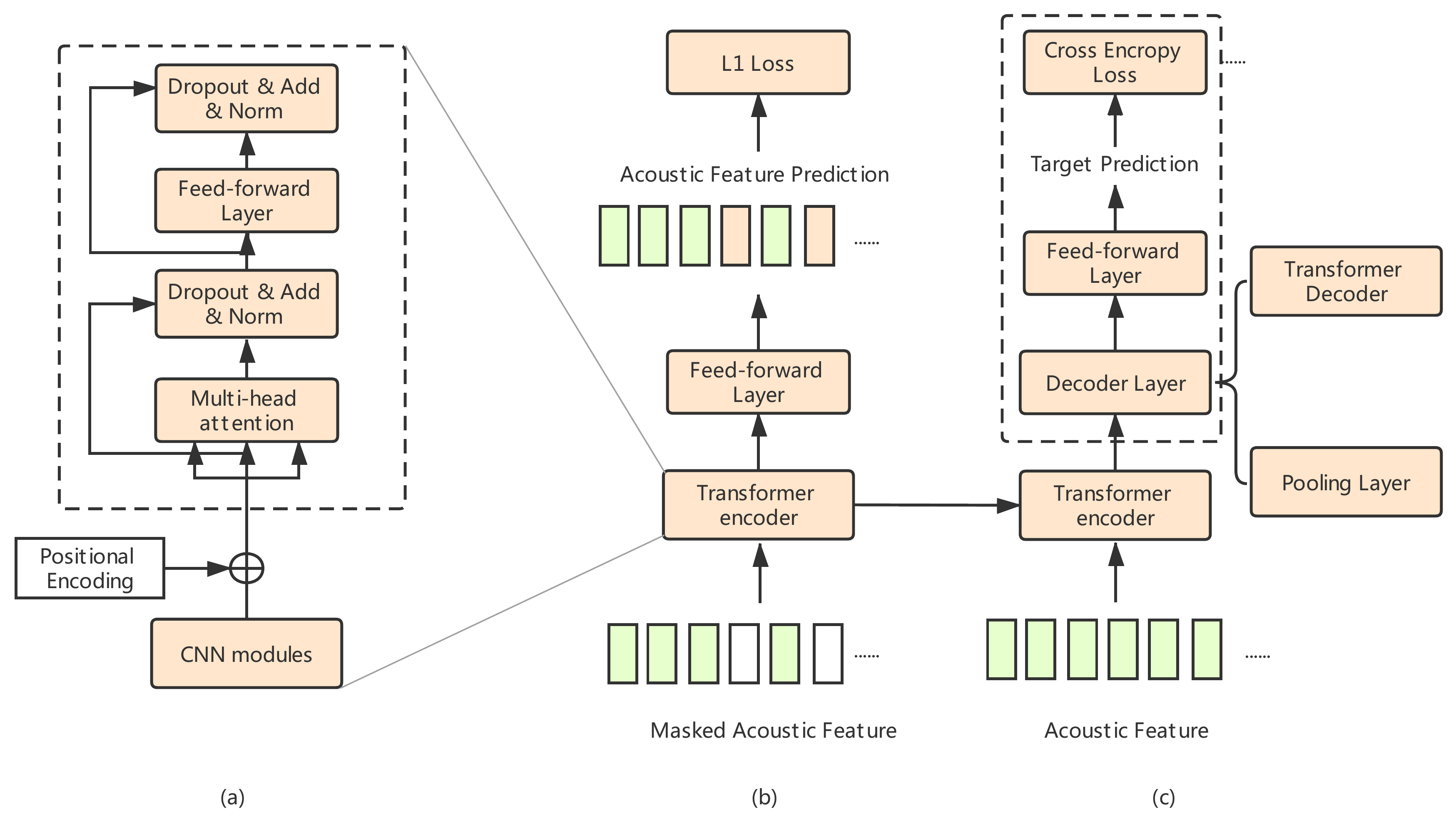}}
		\caption{training structure and procedure: (a) The structure of Transformer based encoder. (b) Pre-training: it is trained to predict the masked acoustic feature using L1 loss. (c) Fine-tuning: the pre-trained transformer encoder is fine-tuned with an additional decoder layer to adapt to the specific task.}\medskip
	\end{minipage}
\end{figure*}

\section{Methodology}
\label{sec:methodology}
To learn a general high-level acoustic representation, we use Transformer based encoder in an unsupervised manner. The architecture of Transformer based encoder is illustrated in Figure 1(a). 

For unsupervised pre-training, Figure 1(b) shows our pre-training procedure. 15\% of frames of the acoustic feature sequence will be masked by zeros and the object of unsupervised pre-training is similar as that of \cite{jiang2019improving} which is to restore the masked frames given the left and right context features. However, we have two aspects that are different from that of \cite{jiang2019improving}. On one hand, we have different masking mechanisms. Generally speaking, the CNN modules of Transformer based encoder provide a downsampling mechanism, by which the frames would be N-fold downsampled. Therefore, to reserve the masked information after downsampling operations, we split frames into chunks each of which contains N frames and 15\% of all chunks will be selected randomly and all frames of the selected chunks will be masked by zeros. On the other hand, Transformer encoder is followed by a feed-forward layer to transform each chunk-level prediction into frame-level predictions. With these changes, we also use L1 loss to minimize the gap between the predicted frames and the corresponding real frames.

For fine-tuning, Transformer encoder needs to be pre-trained only once and can be adapted to varieties of acoustic tasks no matter whether the downstream task deal with the speech or non-speech acoustic sequences, and no matter whether the output of the task is a sequence or a tag. All we should do is to add a decoder layer after the pre-trained encoder to fine-tune the whole model for specific tasks. The choice of decoder layers is based on the tasks as shown in Figure 1(c). We can use Transformer decoder for seq-to-seq tasks and specific pooling layers for tagging tasks. 
\section{Experiments}
\label{sec:experiments}
To prove the effectiveness of our unsupervised pre-training method on various kinds of acoustic tasks, we selected three representative kinds of tasks: speech emotion recognition, acoustic event detection and speech translation. 
\subsection{Data}
To pre-train the model using a larger dataset which can be adapted to various kinds of downstream tasks, we merge MuST-C En-De\cite{di2019must}(408 hours), Librispeech\cite{panayotov2015librispeech}(960 hours) and ESC-US\cite{piczak2015esc}(~347 hours) datasets into one dataset(almost 1715 hours) and we call it OpenAudio. Among them, ESC is an open dataset for environmental sound classification while ESC-US is a compilation of 250k unlabeled clips which were extracted from public field recordings. MuST-C is a multilingual corpus for speech translation from English into 8 languages. For each target language, MuST-C comprises at least 385 hours of audio recordings from English TED Talks. LibriSpeech is a corpus of reading English speech with sampling rate of 16 kHz. The data has been carefully segmented and aligned.

 For pre-training, we did not use speed perturbation but for fine-tuning in every downstream task, we used speed perturbation with factor of 0.9 and 1.1 for data augmentation. We use 40-dimensional Mel filter-banks extracted from the audio signals using window size of 25 ms and step size of 10 ms for pre-training and fine-tuning in all downstream tasks. 

\subsection{Experimental setups}
For Transformer based model, we use the structure discussed before with hidden dimension size of 256, feed-forward size of 2048, attention heads of 4, dropout rate of 0.1 and encoder layers of 12 for all tasks. 

We pre-trained our model using OpenAudio only once and fine-tuned it in each downstream task. It was trained on 4 GPUs with a total batch size of 256 for 50 epochs. We used the Adam optimizer\cite{kingma2014adam} with warmup schedule\cite{vaswani2017attention} according to the formula:
\begin{equation}
	lrate = k * d_{model}^{0.5} * min(n^{-0.5}, n*warmup\_n^{-1.5})
\end{equation}
where n is the step number. k = 0.5 and warmup n = 8000 were chosen for all experiments. For comparison, we also pre-trained our model on each task using its own training data with the same setups as discussed before.
\subsection{Speech emotion recognition}
\label{sec:Speech emotion recognition}

The IEMOCAP database\cite{busso2008iemocap} was commonly used in previous speech emotion studies\cite{pappagari2020x}.
We also use it for our experiments. We used the recordings where majority of annotators agreed on the emotion labels and it contains 4 kinds of emotions: angry, happy, sad and neutral state. Happy and excited emotions were combined as happy in order to balance the number of samples in each emotion class. The dataset contains 5,531 utterances (1,103 angry, 1,636 happy, 1,708 neutral, 1,084 sad) grouped into 5 sessions. We conducted 5-fold cross validation on IEMOCAP, taking samples from 8 speakers for training and the others for evaluation. For fine-tuning, we add an average pooling layer followed by one feed-forward layer. It was trained on 4 GPUs with a total batch size of 64 for 25 epochs. We also use the optimizer which is the same as that of pre-training. For evaluating the performance, we restore the checkpoint averaged from best 5 checkpoints during training. We used UAR which is defined as the unweighted average of the class-specific recalls achieved by the system as our metrics.

In our experiments as shown in Table \ref{tab:Results of speech emotion recognition}, we achieve a mean UAR of 64.9\% which is significantly better than the state-of-the-art result on this setup. According to \cite{Michael2019Improving} and the best of our knowledge,  \cite{Rozgic2012Ensemble} and \cite{xia2015leveraging} presented the best results in the condition that almost match our setups. Specifically, they all use 4 emotion classes and merge happy and excited as one class, except that they used leave-one-speaker-out cross validation and we use leave-one-session-out cross validation. Compared with \cite{Michael2019Improving} which has provided another unsupervised pre-training method, our Transformer based model with pre-training can achieve better performance.

\begin{table}[h]
	\caption{Results of speech emotion recognition (Note: Method and Data represent pre-training method and pre-training data respectively)}
	\label{tab:Results of speech emotion recognition}
	\setlength{\tabcolsep}{3pt} 
	\centering
	\begin{tabular}{cccc}
		\hline
		& \textbf{Method}      & \textbf{Data}    &\textbf{IEMOCAP}     \\
		\hline
		Rozgic et al.\cite{Rozgic2012Ensemble} & - & - & 60.9 \\ 
		Xia et al.\cite{xia2015leveraging} & - & - & 62.5 \\
		Michael et al.\cite{Michael2019Improving} & Autoencoder & TED-LIUM & 59.5 \\
		\hline
		Transformer & - & - & 60.6 \\
		Transformer & Ours & IEMOCAP &  61.8\\
		Transformer & Ours & OpenAudio & {\bfseries 64.9} \\
		\hline
	\end{tabular}
\end{table}

\subsection{Sound event detection}
\label{sec:Sound event detection}
We used DCASE2018 task5 dataset\cite{Dekkers2017} for sound event detection. It contains a continuous recording of one person living in a vacation home over a period of one week. The continuous recordings were split into audio segments of 10s and each segment represents one activity. The dataset presents 10 kinds of activities like cooking, eating and so on. The DCASE2018 task5 has provided development and evaluation datasets for evaluation and test. We use the macro-averaged F1-score as the metrics of this task. It was trained on 4 GPUs with a total batch size of 128 for 50 epochs. We also use the optimizer which is the same as that of pre-training except that k = 0.3. For evaluating the performance, we restore the checkpoint averaged from best 5 checkpoints during training. Similar to speech emotion recognition, we used an average pooling layer as the decoder layer for finetuning.

We compared our work with top three teams' technical reports\cite{Inoue2018, Liu2018, Liao2018} listed on the DCASE community website. Table \ref{tab:Results of sound event detection} shows that with pre-training using OpenAudio, Transformer based model can achieve better performance than all of them on the development set and one of them on the evaluation set. Consider that they used well-designed hand-crafted features with various kinds of data augmentation and ensemble tricks, our method presents a simple but effective training scheme.

\begin{table}[h]
	\caption{Results of sound event detection (Note: Method and Data represent pre-training method and pre-training data respectively, DCASE represets DCASE2018 task5 dataset)}
	\label{tab:Results of sound event detection}
	\setlength{\tabcolsep}{5pt} 
	\centering
	\begin{tabular}{lllll}
		\hline
		& \textbf{Method}      & \textbf{Data}    &\textbf{Dev.}   &\textbf{Eval.}  \\
		\hline
		Inoue et al.\cite{Inoue2018} & - & - & 90.0 & 88.4 \\ 
		Liu et al.\cite{Liu2018} & - & - & 89.8 & 87.5 \\
		Liao et al.\cite{Liao2018} & - & - & 89.8 & 86.7 \\ 
		\hline
		Transformer & - & - & 89.5 & 85.4 \\
		Transformer & Ours & DCASE & 90.4 & 86.6 \\ 
		Transformer & Ours & OpenAudio & {\bfseries 91.0} & {\bfseries 87.5} \\
		\hline
	\end{tabular}
\end{table}

\subsection{Speech translation}
\label{sec:Speech translation}
The aim of speech translation is to translate one language directly from the speech into another language. We used MuST-C English-to-German(En-De) and English-to-French(En-Fr) datasets\cite{di2019must} which were commonly used in previous speech translation studies\cite{Mattia2019Adapting, inaguma2020espnet}. For fine-tuning, we used a 6-layer Transformer decoder as the decoder layer. To avoid overfitting, we also used label smoothing with the rate of 0.1. Similar to \cite{inaguma2020espnet}, we used 8k vocabularies based on byte pair encoding (BPE)\cite{sennrich-etal-2016-neural}. It was trained on 4 GPUs with a total batch size of 512 for 50 epochs. We also use the optimizer which is the same as that of pre-training except that k = 2.5 and warmup n = 25000. For evaluating the performance, we restore the checkpoint averaged from best 5 checkpoints during training. We used beam search with beam size of 10 and performance was evaluated using case-sensitive 4-gram BLEU\cite{papineni2002bleu} on the tst-COMMON set. 

According to \cite{inaguma2020espnet} and \cite{Mattia2019Adapting}, for end-to-end speech translation, Transformer based model has provided state-of-the-art results on MuST-C datasets. However, its performance depends on ASR pre-training which needs English transcripts. In our experiments as shown in Table \ref{tab:Results of speech translation}, we do not need English transcripts and the performance of Transformer pre-trained by its own training audio can be comparable with that of Transformer pre-trained by ASR. Furthermore, because we can easily extend our pre-training data without any specific label, we evaluated the results of Transformer pre-trained by OpenAudio. The results have shown that the BLEU scores have exceeded that of \cite{inaguma2020espnet} pre-trained by ASR on both datasets.

We can see that different from current end-to-end speech translation methods, our methods provide not only better performance but an easier training scheme without transcripts of speech in same language which is more practical for industrial application. It is also promising that combining our unsupervised pre-training method with the current supervised pre-training mechanism will further improve the performance.
\begin{table}[h]
	\caption{Results of speech translation (Note: Method and Data represent pre-training method and pre-training data respectively)}
	\label{tab:Results of speech translation}
	\setlength{\tabcolsep}{5pt} 
	\centering
	\begin{tabular}{ccccc}
		\hline
		& \textbf{Method}      & \textbf{Data}    &\textbf{En-De}    &\textbf{En-Fr}       \\
		\hline
		Pipeline\cite{Mattia2019Adapting} & - & - & 18.50 & 27.90 \\ 
		Transformer\cite{inaguma2020espnet} & - & - & 16.40 & N/A \\
		Transformer\cite{inaguma2020espnet} & ASR & MuST-C & 21.77 & 31.56 \\ 
		\hline
		Transformer & - & - & 19.64 & 29.40  \\
		Transformer & ASR & MuST-C & 21.93 & 31.70 \\ 
		Transformer & Ours & MuST-C & 21.50 & 31.32 \\ 
		Transformer & Ours & OpenAudio & {\bfseries 22.04} &  {\bfseries 31.88} \\
		\hline
	\end{tabular}
\end{table}

\section{Conclusion}
\label{sec:Discussions And Conclusion}
In this work, we explored Transformer based encoder with Masked-LM like pre-training for acoustic representation learning. We conducted experiments on three kinds of tasks: speech emotion recognition, sound event detection and speech translation. We pre-train the model with a large dataset combining Librispeech, MuST-C and ESC-US datasets and fine-tune it on each task. Results have shown that for speech translation, the BLEU score can improve relatively 12.2\% and 8.4\% on MuST-C En-De and En-Fr datasets respectively compared with that of Transformer without pre-training. For sound event detection, the F1 score can improve absolutely 1.5\% and 2.1\% on DCASE2018 task5 development set and evaluation set compared with that of our base Transformer. For speech emotion recognition, the UAR can improve absolutely 4.3\% on IEMOCAP dataset compared with that of our base Transformer. 

Compared with current state-of-the-art acoustic systems, our method is able to provide a more general and robust acoustic representation for all acoustic tasks and it is easy to be transferred, easy to be built without many hand-crafted designs and is more practical for industrial applications. It suggests that our method can provide a promising alternative for acoustic representation learning.

\bibliographystyle{IEEEtran}
\bibliography{refs}

\end{document}